\documentclass[conference,letter]{IEEEtran}

\usepackage{amsmath}
\usepackage{amsfonts}
\usepackage{amssymb}
\usepackage{epsf}
\usepackage{cite}
\usepackage{balance}
\usepackage{fancyhdr}
\usepackage{graphicx}
\usepackage{subfigure}
\usepackage[blocks]{authblk}
\usepackage[stable]{footmisc}
\usepackage{textcase}
\usepackage{hyperref}
\usepackage{fancyhdr}

\makeatletter
\def\ps@IEEEtitlepagestyle{%
  \def\@oddfoot{\mycopyrightnotice}%
  \def\@evenfoot{}%
}
\def\mycopyrightnotice{%
  {\footnotesize \copyright 2015 IEEE. Personal use of this material is permitted. Permission from IEEE must be obtained for all other uses, in any current or future media\hfill}
  \gdef\mycopyrightnotice{}
}

\newcounter{subeq}

\IEEEoverridecommandlockouts
\date{}
\begin{document}
\title{On the Performance of Single- and Multi-carrier Modulation Schemes for Indoor Visible Light Communication Systems}

\author{Mohammadreza A.Kashani}
\author{Mohsen Kavehrad}

\affil{Department of Electrical Engineering\authorcr
Pennsylvania State University, University Park, PA 16802\authorcr
Email: \{mza159, mkavehrad\}@psu.edu\authorcr}

\maketitle
\begin{abstract}
In this paper, we investigate and compare the performance of single- and multi-carrier modulation schemes for indoor visible light communication (VLC). Particularly, the performances of single carrier frequency domain equalization (SCFDE), orthogonal frequency division multiplexing (OFDM) and on-off keying (OOK) with minimum mean square error equalization (MMSE) are analyzed in order to mitigate the effect of multipath distortion of the indoor optical channel where nonlinearity distortion of light emitting diode (LED) transfer function is taken into account. Our results indicate that SCFDE system, in contrast to OFDM system, does not suffer from high peak to average power ratio (PAPR) and can outperform OFDM and OOK systems. We further investigate the impact of LED bias point on the performance of OFDM systems and show that biasing LED with the optimum value can significantly enhance the performance of the system. Bit-interleaved coded modulation (BICM) is also considered for OFDM and SCFDE systems to further compensate signal degradation due to inter-symbol interference (ISI) and LED nonlinearity. \end{abstract}

\section{Introduction}\label{INTRODUCTION}
\IEEEPARstart{V}isible light communication (VLC) refers to unguided optical transmission via the use of light emitting diodes (LEDs) \cite{komine2004fundamental,randel2010advanced,street1997indoor}. Indoor VLC is characterized by short transmission range and free from major outdoor environmental degradations such as rain, snow, building sway, and atmospheric turbulence. This technology has recently attracted significant attentions as a promising complementary technology for radio frequency (RF) in short-range communications \cite{randel2010advanced,komine2003integrated,rajagopal2012ieee,kavehrad2007broadband}. These systems offer signiﬁcant technical and operational advantages such as higher bandwidth capacity, virtually unlimited reuse, unregulated spectrum and robustness to electromagnetic interference. Despite the major advantages of indoor VLC, they suffer from multipath distortion due to dispersion of the optical signal caused by reflections from various sources inside a room. This dispersion leads to inter-symbol interference (ISI) at high data rates which reduces signal to noise ratio (SNR) and severely impairs the link performance.

Multicarrier modulation which is usually implemented by orthogonal frequency division multiplexing (OFDM) has been originally introduced in RF communication to mitigate ISI and multipath dispersion. The concept of OFDM has been applied to indoor wireless optical communications (WOC) in \cite{shieh2008coherent,gonzalez2005ofdm,elgala2009indoor,armstrong2006power,armstrong2009ofdm,carruthers1996multiple} to support high data rates. However, the performance of VLC systems using intensity-modulation direct-detection (IM/DD) along with OFDM modulation is significantly affected by nonlinear characteristic of LED due to the large peak-to-average power ratio (PAPR) of OFDM signal. In particular, signal amplitudes below the LED turn-on-voltage (TOV) and above the LED saturation point are clipped. To address this issue, single-carrier frequency domain equalization (SCFDE) has been proposed in the literature reducing PAPR while achieving similar throughput as OFDM systems \cite{ciochina2010review,falconer2002frequency}. Several OFDM schemes such as DC-clipped OFDM \cite{kahn1997wireless}, asymmetrically clipped optical OFDM (ACO-OFDM) \cite{armstrong2006power} and PAM-modulated discrete multitone (PAM-DMT) \cite{lee2009pam} have been proposed in the literature. Among these schemes, ACO-OFDM has been shown to be more efficient in terms of optical power than the systems that use DC-biasing as it utilizes a large dynamic range of the LED. Therefore, it is considered in this paper.

There exist several investigations analyzing different OFDM techniques and comparing them with SCFDE \cite{mesleh2012ofdm,mesleh2011performance,dissanayake2013comparison} or single carrier modulation \cite{barros2012comparison,armstrong2008comparison}. To the best of our knowledge, these previous studies were built on the assumption of ideal additive white Gaussian noise (AWGN) channels or did not consider the nonlinear characteristics of LED. In this paper, we analyze and compare performance of the aforementioned techniques along with on-off keying (OOK) with minimum mean square error equalization (MMSE) which is commonly used in IM/DD communication systems considering an off-the-shelf LED model and a multipath channel. Moreover, bit-interleaved coded modulation (BICM) is considered for OFDM and SCFDE systems to further combat signal degradation due to LED nonlinearity and ISI.

\begin{figure*}[t]
\centering
\includegraphics[width = 13cm, height = 6cm]{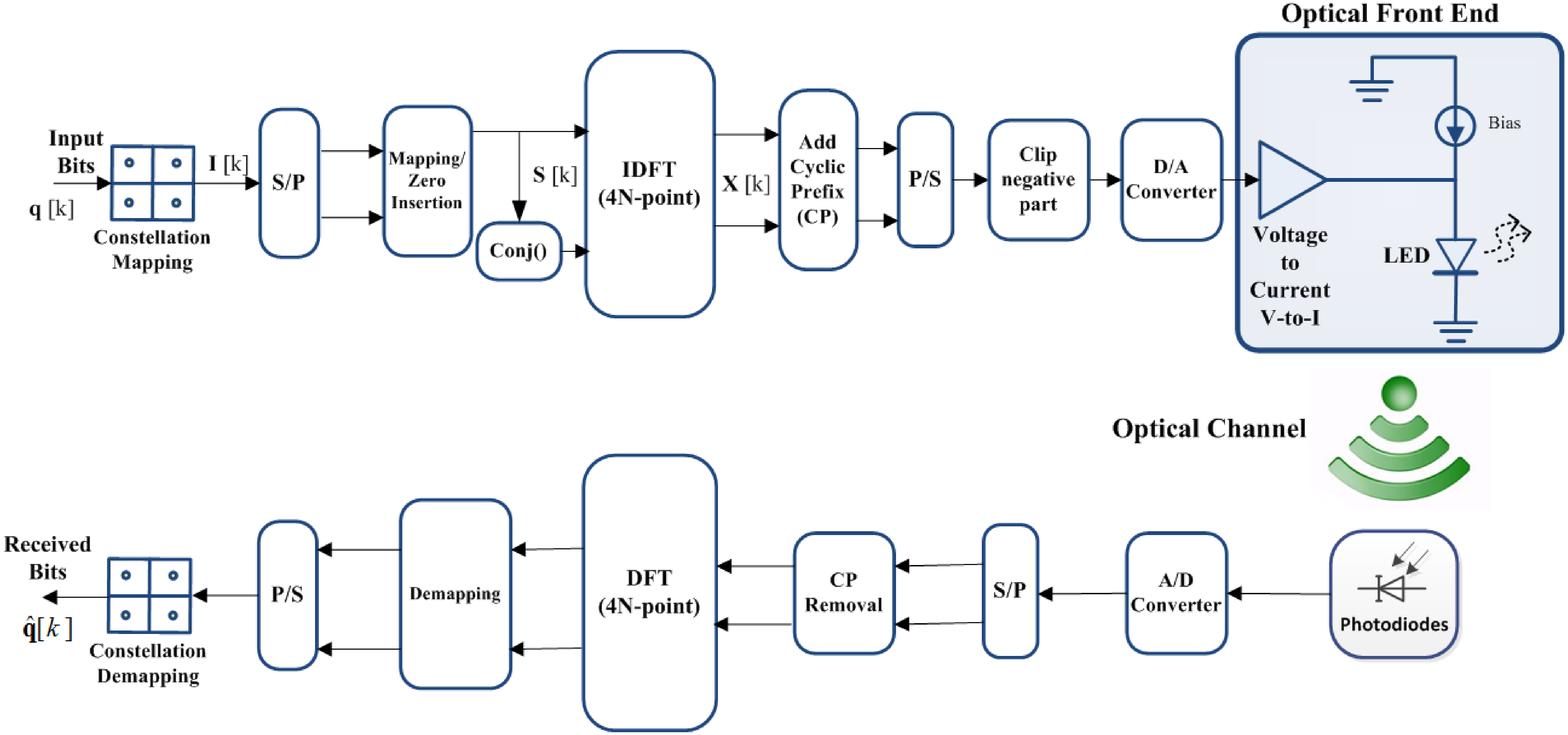}
\caption{ACO-OFDM Transmitter and Receiver configuration.}
\label{aco}
\end{figure*}
The rest of the paper is organized, as follows. In Sections II and III, we briefly introduce the system model and describe ACO-OFDM and ACO-SCFDE.  In Section IV, we compare the PAPR performance of ACO-OFDM and ACO-SCFDE. In Section V, we analyze and compare ACO-SCFDE, ACO-OFDM and OOK performance and investigate the impact of LED bias point on the performance. Furthermore, we show that BICM can combat signal degradation due to LED nonlinearity and ISI. Finally, Section VI concludes the paper.
\section{System Model of Asymmetrically-Clipped Optical OFDM (ACO-OFDM)}

ACO-OFDM is a form of OFDM that modulates the intensity of a LED. Because ACO-OFDM modulation employs IM/DD, the time-domain transmitted signal must be real and positive. The block diagram of an ACO-OFDM system is depicted in Fig. \ref{aco}. The information stream is first parsed into a block of $N$ complex data symbols denoted by $\mathbf{I}={{[{{I}_{0}},{{I}_{1}},\ldots ,{{I}_{N-1}}]}^{T}}$ where the symbols are drawn from constellations such as $M$-QAM or $M$-PSK where $M$ is the constellation size. To ensure a real output signal used to modulate the LED intensity, ACO-OFDM subcarriers must have Hermitian symmetry. In ACO-OFDM, only odd subcarriers are modulated, and this results in avoiding the impairment from clipping noise. Therefore, the complex symbols are mapped onto a $4N\times 1$ vector as $\mathbf{S}={{[0,{{I}_{0}},0,{{I}_{1}},\ldots ,0,{{I}_{N-1}},0,I_{N-1}^{*},0,\ldots ,I_{1}^{*},0,I_{0}^{*},0]}^{T}}$ where ${\left( . \right)}^{*}$ denotes the complex conjugate of a vector. A $4N$-point IFFT is then applied on the vector $\mathbf{S}$ to build the time domain signal $\mathbf{x}$. A cyclic prefix (CP) is added to $\mathbf{x}$ turning the linear convolution with the channel into a circular one to mitigate multipath dispersion. To make the transmitted signal unipolar, all the negative values are clipped to zero. It is proven in \cite{armstrong2006power} that since only the odd subcarriers are used to carry the data symbols, the clipping does not affect the data-carrying subcarriers, but only reduces their amplitude by a factor of two.

The unipolar signal is then converted to analog and filtered to modulate the intensity of an LED. At the receiver, the signal is converted back to digital. CP is then removed and the electrical OFDM signal is demodulated by taking a $4N$ FFT and equalized with a single-tap equalizer on each subcarrier to compensate for
channel distortion.  The even subcarriers are then discarded and the transmitted data is recovered by a hard or soft decision. The extraction of odd subcarriers along with the equalization are represented by the \emph{Demapping} block in Fig. \ref{aco}.
\section{System Model of Asymmetrically-Clipped Optical SCFDE (ACO-SCFDE)}
\begin{figure*}[t]
\centering
\includegraphics[width = 13cm, height = 6cm]{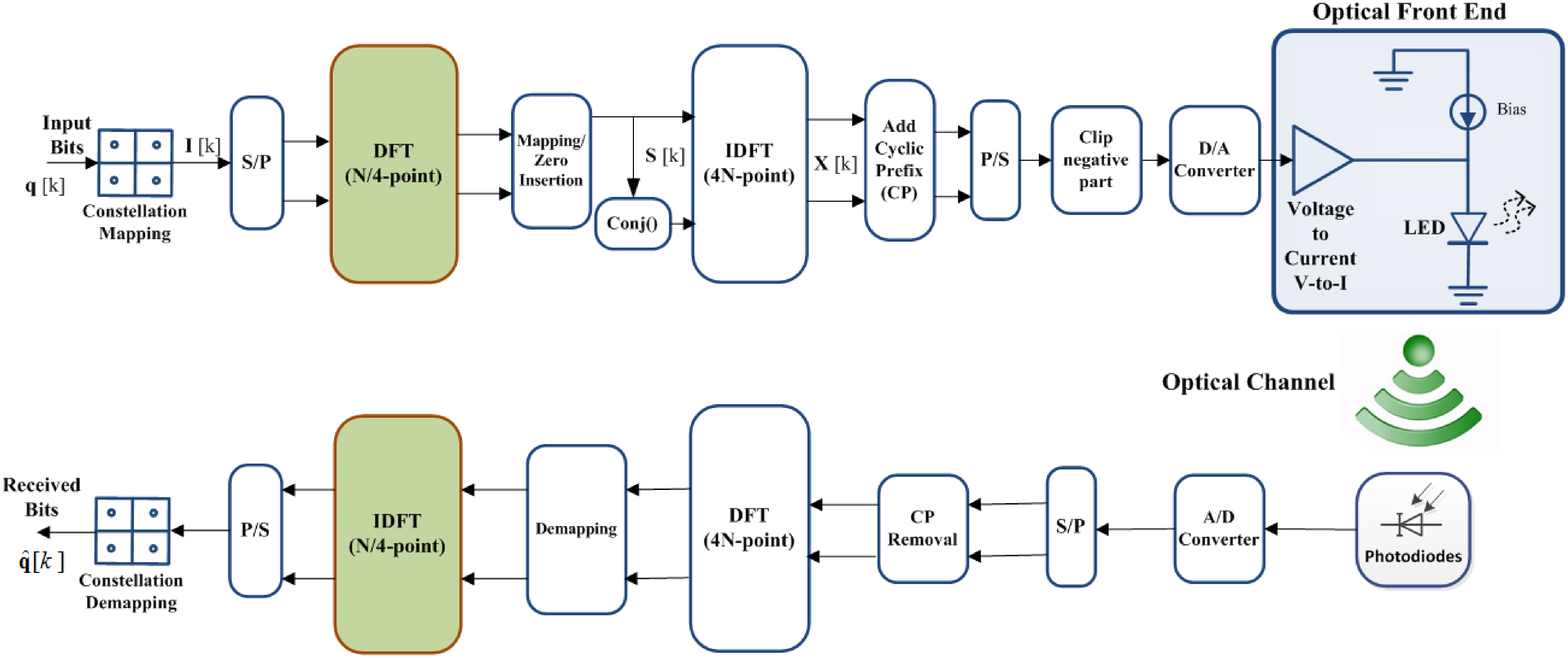}
\caption{ACO-SCFDE Transmitter and Receiver configuration.}
\label{sc}
\end{figure*}

SCFDE is a special technique which is compatible with any of OFDM techniques. In this paper, we apply asymmetrically-clipped optical to SCFDE to achieve ACO-SCFDE with low PAPR. The block diagram of an ACO-SCFDE system is depicted in Fig. \ref{sc}. ACO-SCFDE and ACO-OFDM are the same except that in ACO-SCFDE, an extra $N$-point FFT and IFFT are used at the transmitter and the receiver respectively resulting in a single carrier transmission instead of multicarrier. As it will be shown latter, the additional complexity of the extra FFT and IFFT blocks is offset by the fact that SCFDE has lower PAPR and better bit-error-rate (BER) performance than its OFDM counterpart when the signal is sent through the non-linear LED.

\section{Peak-to- Average Power Ratio (PAPR)}
\begin{figure}
\centering
\includegraphics[width = 8cm, height = 7.5cm]{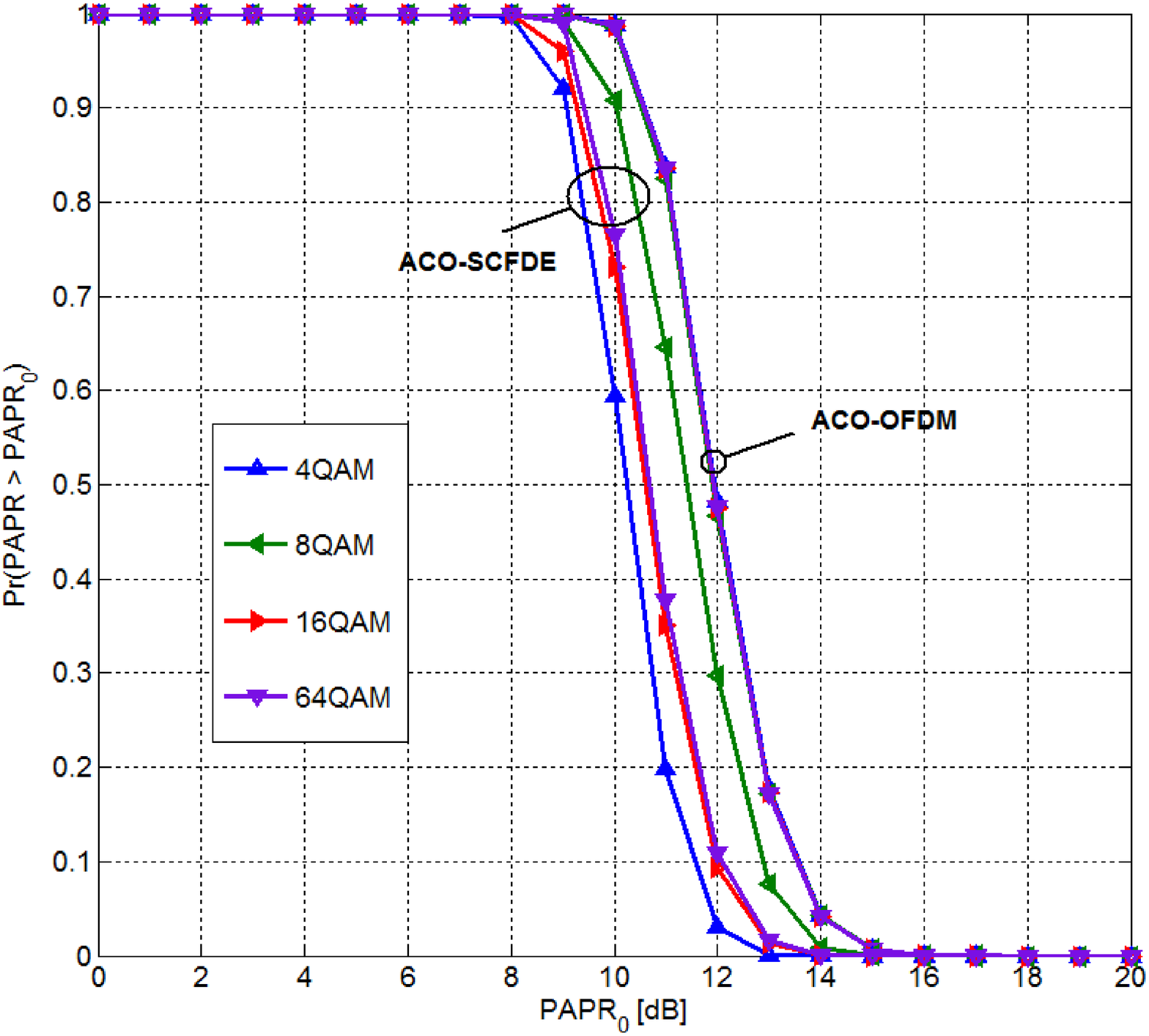}
\caption{CCDF of PAPR Comparison of ACO-OFDM and ACO-SCFDE for $N=64$.}
\label{p64}
\end{figure}
\begin{figure}
\centering
\includegraphics[width = 8cm, height = 7.5cm]{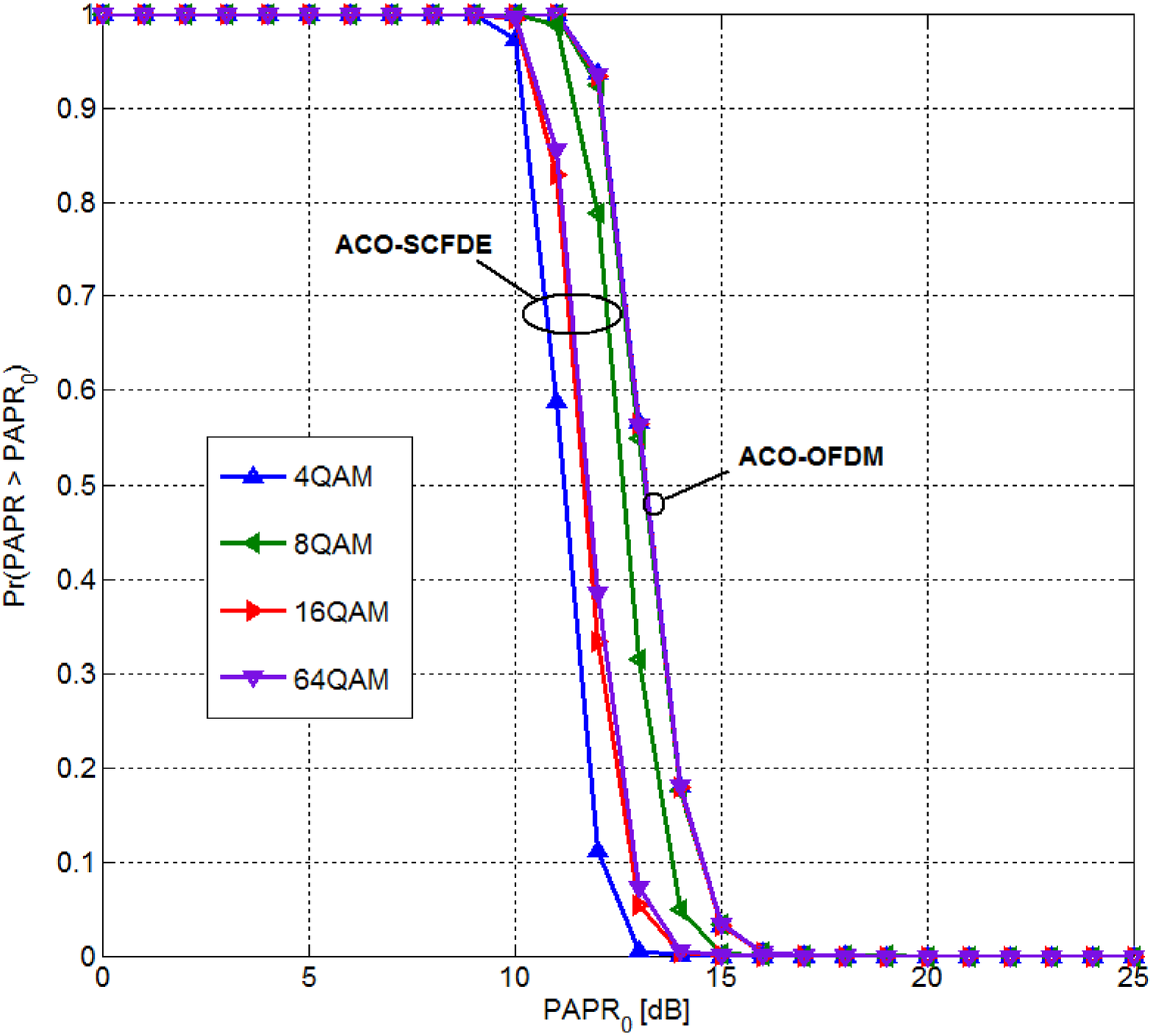}
\caption{CCDF of PAPR Comparison of ACO-OFDM and ACO-SCFDE for $N=256$.}
\label{p256}
\end{figure}

In this section, the PAPR of ACO-OFDM signals is analyzed and compared with that of ACO-SCFDE. The PAPR is defined as the maximum power of transmitted signal divided by the average power, that is
\begin{equation}
PAPR=\frac{\max {{x}^{2}}\left( n \right)}{E\left[ {{x}^{2}}\left( n \right) \right]}
\end{equation}
where $E\left[.\right]$ denotes expectation. Due to the large number of subcarriers and occasional constructive combining of them, OFDM systems have a large dynamic signal range and exhibit a very high PAPR.
Thus, the OFDM signal will be clipped when passed through a nonlinear LED at the transmitter end which results in degrading the BER performance. SCFDE can be used as a promising alternative technology for OFDM to reduce the PAPR and combat the effect of nonlinear characteristics of the LED.

 PAPR is usually presented in terms of a complementary cumulative distribution function (CCDF) which is the probability that PAPR is higher than a certain PAPR value $PAP{{R}_{0}}$, i.e. $\Pr \left( PAPR>PAP{{R}_{0}} \right)$. Figs \ref{p64}-\ref{p256} demonstrate the CCDF of PAPR for $N=64$ and 256 subcarriers respectively, calculated by Monte Carlo simulation for different modulation constellations. We notice that ACO-SCFDE has a lower PAPR as compared to ACO-OFDM system for the same number of subcarriers. We also observe that the PAPR increases with increasing $N$ for all of the constellations.

\section{Performance Analysis}

\begin{figure}
\centering
\includegraphics[width = 7.5cm, height = 6cm]{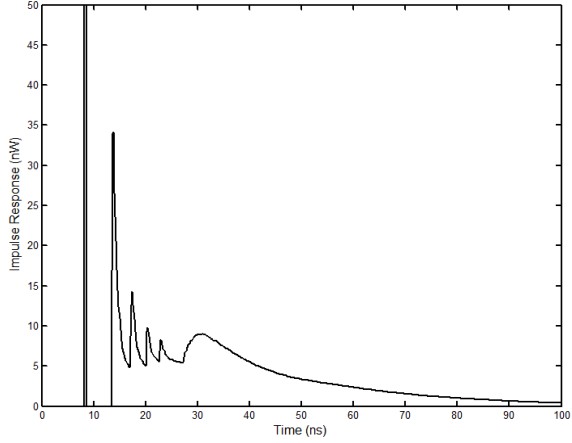}
\caption{Impulse response of the indoor diffuse channel.}
\label{channel}
\end{figure}

\begin{table}

\begin{center}
\begin{quote}
\caption{Room configuration under consideration.}
\label{table}
\end{quote}
{\small{
\begin{tabular}{|c|c|c|}
  \hline
  Room &Length  &6 m  \\ \cline{2-3}
  &Width &5 m \\ \cline{2-3}
  &Height &3 m\\ \hline
  Reflectivity &${\rho }_{\text{North}}$ &0.8\\ \cline{2-3}
                &${\rho }_{\text{South}}$ &0.8\\ \cline{2-3}
                &${\rho }_{\text{East}}$ &0.8\\ \cline{2-3}
                &${\rho }_{\text{West}}$ &0.8\\ \cline{2-3}
                &${\rho }_{\text{Ceiling}}$ &0.8\\ \cline{2-3}
                &${\rho }_{\text{Floor}}$ &0.3\\ \hline
  Source &Mode &1 \\ \cline{2-3}
        &Azimuth &${{0}^{\circ }}$ \\ \cline{2-3}
        &Elevation &${{-90}^{\circ }}$ \\ \cline{2-3}
        &$x$, $y$, $z$ &0.1 m, 0.2 m, 3 m\\ \hline
  Receiver &Area &$\text{CM}^{2}$\\ \cline{2-3}
            &FOV &${{0}^{\circ }}$ \\ \cline{2-3}
            &$x$, $y$, $z$ &2.5 m, 2.5 m, 1 m\\ \hline
\end{tabular}}}
\end{center}
\end{table}
 Simulations are conducted assuming indoor optical multipath channel where the transmitter and receiver are placed in a room whose configuration is summarized in Table \ref{table}. The methodology developed by Barry et al \cite{barry1993simulation} is employed to simulate the impulse response of the channel where 10 reflections are taken into account. Fig. \ref{channel} presents the impulse respond of a diffuse channel.

\begin{figure} \centering
\subfigure[]{\includegraphics[width = 7cm, height = 5cm]{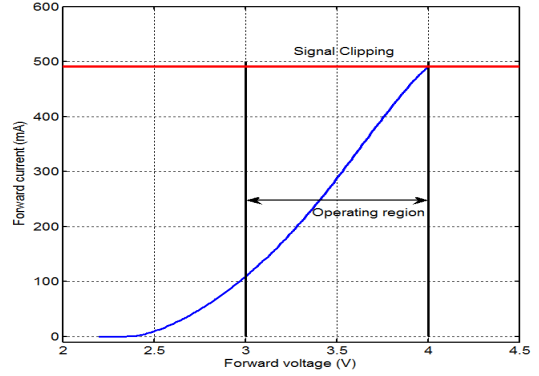}}
\subfigure[]{\includegraphics[width = 7cm, height = 5cm]{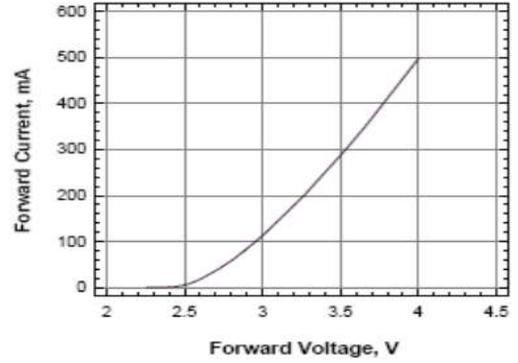}}
\caption{Transfer characteristics of OPTEK, OVSPxBCR4 1-Watt white LED. (a) Fifth-order polynomial fit to the data. (b) The curve from the data sheet.}
\label{led}
\end{figure}

We assume an OFDM signal whose average electrical power before modulating the LED is varied from -10 dBm to 30 dBm, and the power of AWGN is -10 dBm. Thus, the simulated electrical SNR ranges from 0 dB to 40 dB matching the reported SNR values for indoor WOC systems \cite{o2007optical,grubor2008bandwidth}. A number of subcarriers of $N=64$ with $M$-QAM modulation are also assumed. Furthermore, OPTEK, OVSPxBCR4 1-Watt white LED is considered in simulations whose optical and electrical characteristics are given in Table \ref{table2}. A polynomial order of five is used to realistically model measured transfer function. Fig. \ref{led} demonstrates the non-linear Transfer characteristics of the LED from the data sheet and using the polynomial function.

\begin{table}
\begin{center}
\begin{quote}
\caption{Optical and Electrical Characteristics of OPTEK, OVSPxBCR4 1-Watt white LED.}
\label{table2}
\end{quote}
{\small{
\begin{tabular}{|c|c|c|c|c|c|}
  \hline
  \textbf{Symbol} &\textbf{Parameter}  &\textbf{MIN} &\textbf{TYP} &\textbf{MAX} &\textbf{Units}\\ \hline
  $V_F$  &Forward Voltage &3.0 &3.5 &4 &$V$\\ \hline
  $\Phi$  &Luminous Flux &67 &90 &113 &lm\\ \hline
  ${{\Theta }^{1/2}}$  &50\% Power Angle &--- &120 &--- &deg\\ \hline
  \end{tabular}}}
\end{center}
\end{table}

\begin{figure}
\centering
\includegraphics[width = 8cm, height = 7.5cm]{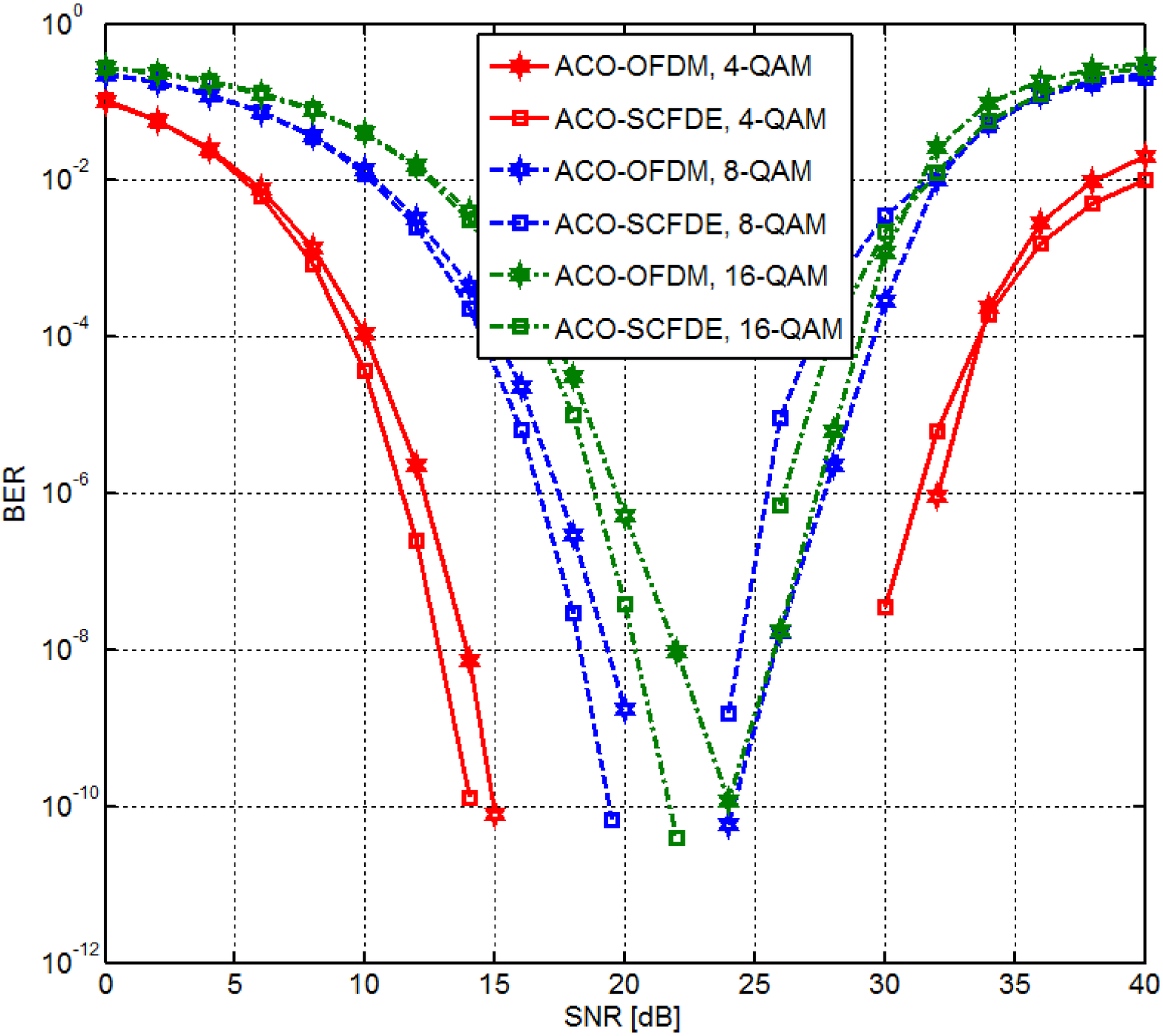}
\caption{BER Comparison of ACO-OFDM and ACO-SCFDE for bias point of 3.2V.}
\label{aco1}
\end{figure}

\begin{figure}
\centering
\includegraphics[width = 8cm, height = 7.5cm]{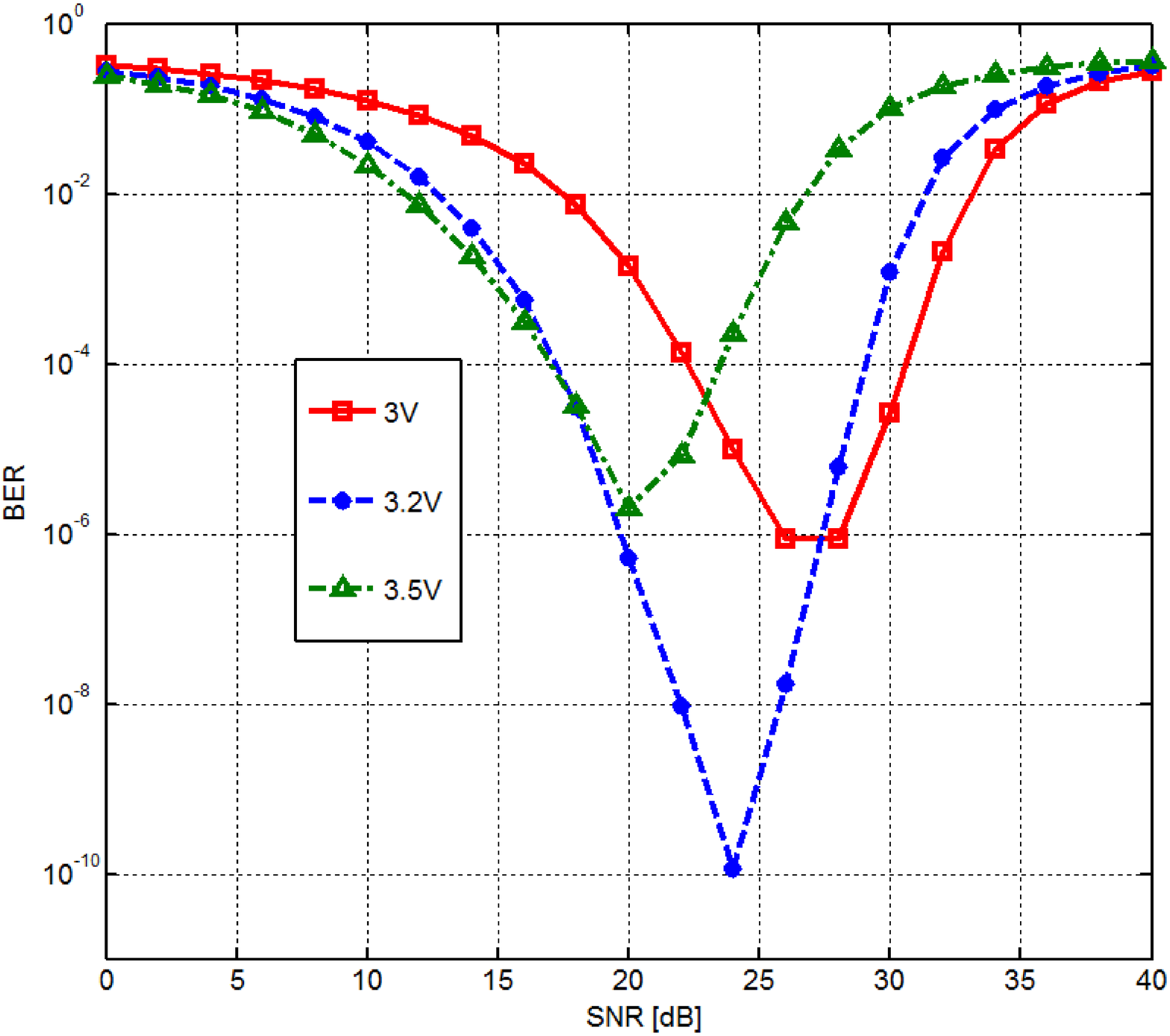}
\caption{BER of ACO-OFDM for $M = 16$ for different bias points.}
\label{bias}
\end{figure}

We first compare the bit error performance of ACO-SCFDE and ACO-OFDM. Fig. \ref{aco1} presents the BER performance of ACO-OFDM and ACO-SCFDE for different modulation orders and LED bias point of 3.2V. As the results indicate, SCFDE exhibits better BER performance in the optical multipath channel.
Furthermore, we investigate the impact of LED bias point on the performance of ACO-OFDM systems. According to the data sheet of the LED used in the simulations, three different bias points (3V, 3.2V and 3.5V) are considered. Fig. \ref{bias} demonstrates BER performance of an ACO-OFDM system with $M=16$ and different LED bias points. As it can be clearly seen, nonlinearity of LED has a significant impact on the performance of optical OFDM systems. It is also observed that there is an optimum LED bias point which is 3.2V for the case under consideration from which deviation can significantly deteriorate the system performance.

\begin{figure}
\centering
\includegraphics[width = 8cm, height = 7.5cm]{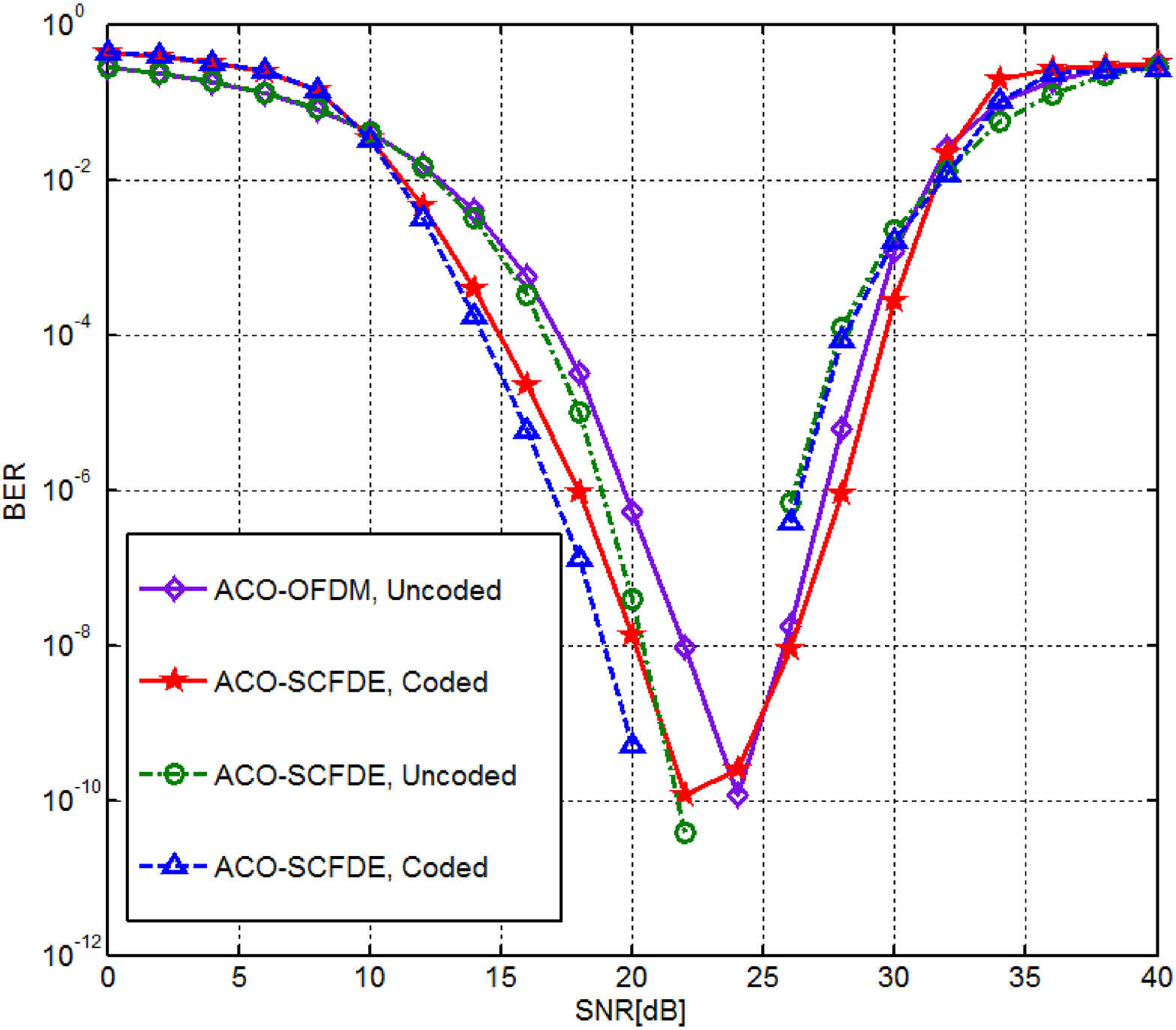}
\caption{BER Comparison of uncoded and coded ACO-OFDM and ACO-SCFDE for $M = 16$.}
\label{code}
\end{figure}

Bit-interleaved coded modulation (BICM) \cite{caire1998bit} is also considered for OFDM and SCFDE systems to further compensate signal degradation due to ISI and LED nonlinearity. To demonstrate the usefulness of BICM, we assume that the information sequence is first encoded by a rate 1/2 convolutional encoder with generator matrix $\mathbf{g} = (5,7)$, constraint length of 3 and minimum Hamming distance of 5. The coded information is then interleaved by a bitwise interleaver. At the receiver, the Viterbi soft-decoder \cite{viterbi1971convolutional} and the de-interleaver are used. Fig. \ref{code} shows the BER of uncoded and coded ACO-OFDM and ACO-SCFDE for the indoor visible light communication under consideration. As it can be clearly observed, BICM can significantly enhance the system performance. However, the achieved gains come at the cost of significant reduction in the data rate due to the insertion of coded bits.

\begin{figure}
\centering
\includegraphics[width = 8cm, height = 7.5cm]{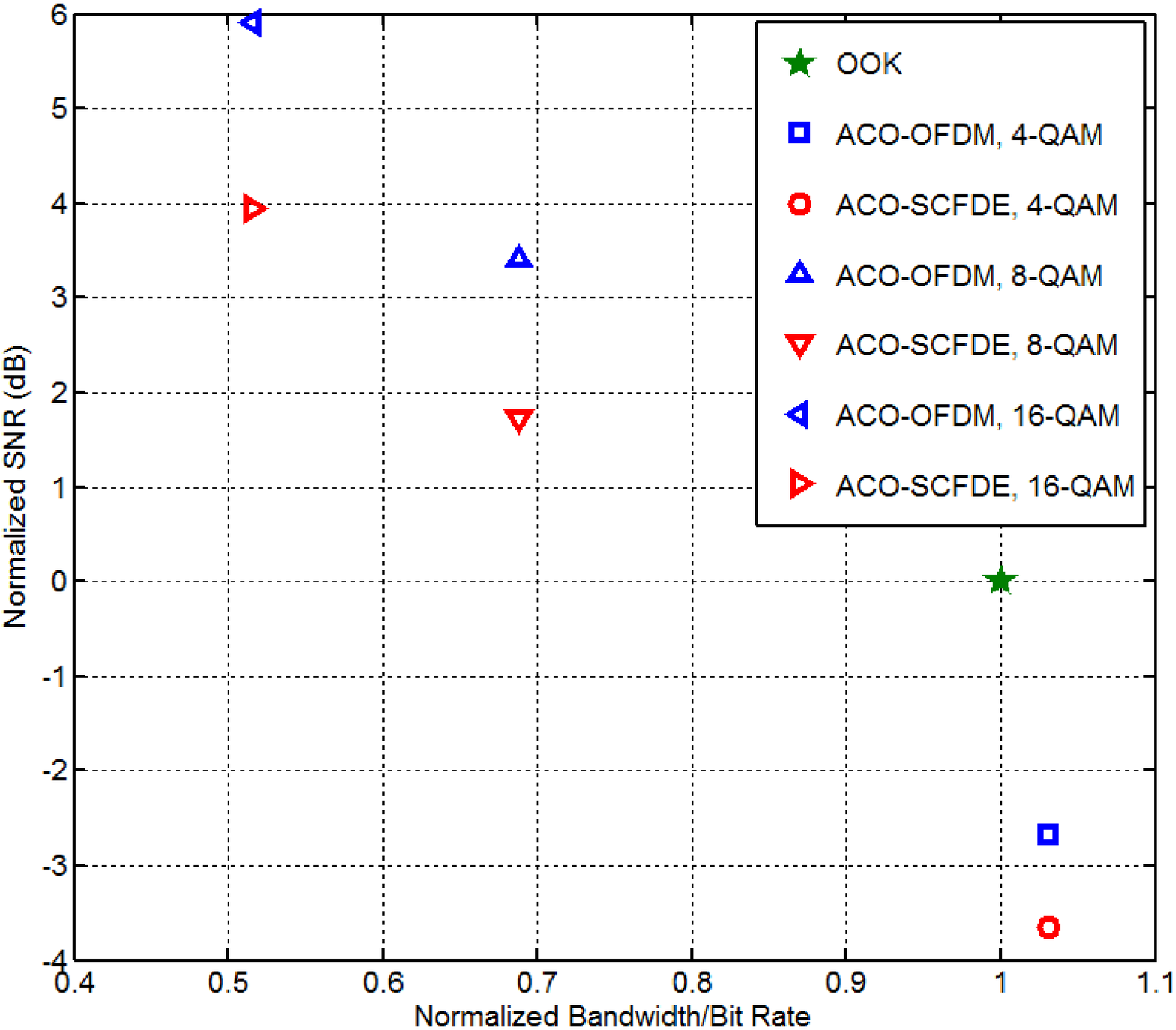}
\caption{Normalized SNR versus normalized bandwidth/bit-rate required to achieve BER of ${{10}^{-9}}$. }
\label{comp}
\end{figure}

Finally, we compare the performance of ACO-OFDM, ACO-SCFDE and OOK modulation with MMSE over an indoor VLC medium. The performance comparison is done in terms of \emph{normalized SNR} and \emph{normalized bandwidth/bit-rate} relative to OOK \cite{kahn1997wireless}. According to \cite{kahn1997wireless} and \cite{armstrong2008comparison}, we define the modulation bandwidth as the position of the first spectral null. To make a fair comparison between different modulation schemes, the normalized bandwidth of the signal is calculated as the modulation bandwidth which is normalized relative to OOK of the same transmitted data rate. For ACO-OFDM and ACO-SCFDE, first null occurs at a normalized frequency of $1+2/N$. Thus, the normalized bandwidth/bit-rate is obtained as $2\left( 1+2/N \right)/\log _{2}^{M}$ for ACO-OFDM and ACO-SCFDE. Fig. \ref{comp} shows normalized SNR required for a BER of ${{10}^{-9}}$ as a function of normalized bandwidth/bit-rate for OOK, ACO-OFDM and ACO-SCFDE. We observe while ACO-OFDM and ACO-SCFDE with 4-QAM modulation of each subcarrier require approximately the same bandwidth as OOK, they are more efficient in terms of power. Particularly, ACO-OFDM and ACO-SCFDE are 2.7 dB and 3.7 dB more efficient than OOK, respectively. For the higher orders of $M$, OOK outperforms ACO-OFDM and ACO-SCFDE but it requires greater bandwidth.

\section{Conclusions}
We have evaluated and compared the performance of IM/DD single- and multi-carrier modulation schemes for indoor visible light communication systems taking into account both nonlinear characteristics of LED and dispersive nature of optical wireless channel. We have shown through the use of simulation that SCFDE system has a lower PAPR than its counterpart OFDM system and outperforms OOK and OFDM systems and therefore is a promising modulation technique for indoor VLC systems. We have also investigated the performance of OFDM systems for different LED bias points and shown that significant gain can be achieved by biasing LED with the optimum value. BICM technique has been further considered to combat signal degradation due to LED nonlinearity and dispersive nature of the channel.
\section{Acknowledgement}
The authors would like to thank the National Science Foundation (NSF) ECCS directorate for their support of this work under Award \# 1201636, as well as Award \# 1160924, on the NSF “Center on Optical Wireless Applications (COWA– \href{http://cowa.psu.edu}{http://cowa.psu.edu})”

\balance
\bibliographystyle{IEEEtran}
\bibliography{ref}
\end{document}